\documentstyle[a4,11pt,epsfig,twoside]{article}
\begin{document}
\setcounter{page}{1}
\title{Determining the Mass of Supersymmetric Scalars at the 
CLIC Multi-TeV $e^+e^-$ Collider}

\author{M. Battaglia
        \thanks{e-mail address: Marco.Battaglia@cern.ch}
        ~and M. Gruw\'e
        \thanks{e-mail address: Magali.Gruwe@cern.ch}
\\
\\
        {\it CERN, Geneva, Switzerland}
}
\date{}
\maketitle
\begin{abstract}
The determination of the smuon mass at the {\sc Clic} multi-TeV $e^+e^-$ linear 
collider has been studied for two CMSSM benchmarks. Results are given for both the 
analysis of the muon energy spectrum and the threshold scan method. 
The effects of detector resolution, beam-beam interactions and accelerator-induced 
backgrounds are discussed. The energy spectrum technique is also applied to the 
$\tilde{t} \to t \tilde{g}$ process to determine the scalar top mass, in scenarios with 
the gluino lighter than the squarks.
\end{abstract}

\section{Introduction}

A multi-TeV $e^+e^-$ linear collider (LC) is expected to complete the investigation of 
the SUSY particle spectrum, in the case Supersymmetry is realised in nature and the 
particle partners exhibit a mass spectrum extending up to 1~TeV, and beyond. 
In this case, precise determinations of the masses and widths of these particles 
will represent a central part of the collider physics programme. 
Sets of SUSY benchmarks have recently been proposed, which take into 
account constraints from direct searches at {\sc Lep-2} and other colliders and data 
from $b \rightarrow s \gamma$ and cosmology~\cite{cmssm,ssp}. Most of these benchmarks 
have the masses of the heaviest sparticles beyond the reach of the {\sc Lhc} and of a 
TeV-class $e^+e^-$ linear collider. The {\sc Lhc} is expected to reveal the 
existence of Supersymmetry and perform a study of the sparticle properties in the vast 
majority of conceivable scenarios and a 0.5-1.0~TeV $e^+e^-$ collider can improve the 
accuracy in the study of the lightest SUSY states. However, higher energies are likely 
to be needed to access all the sparticles species and acquire the data necessary to 
clarify the nature of Supersymmetry breaking. 

A first mapping of the potential of a 3-5~TeV $e^+e^-$ collider, such as 
{\sc Clic}~\cite{clic}, in comparison to the {\sc Lhc} and to a lower energy $e^+e^-$ 
collider was performed, based on a simple assumption on the minimum observable 
product of production cross section and decay branching fraction into well 
reconstructable final states~\cite{cmssm}. Beyond detection, it is 
important to assess the ability of {\sc Clic} to perform accurate determinations of 
sparticle masses, widths and decay branching fractions, taking into account the details 
of beam-beam interactions and the resulting luminosity spectrum and accelerator-induced 
backgrounds~\cite{guinea_pig}. 
 We study here the expected {\sc Clic} accuracy on sfermion masses, using a
realistic simulation. Results are presented for the Supersymmetric partner of the 
muon, $\tilde{\mu_L}$ and for the scalar top, $\tilde{t_2}$. 
Section~2 presents the benchmarks considered and the {\sc Clic} parameters. There are 
two different techniques for the determination of the smuon mass: the analysis of the 
energy distribution of the resulting muon in the two body $\tilde{\mu_L}$ decay and the 
energy scan close the the production threshold. These are discussed in 3.1 and 3.2 
respectively. In Section~4 we discuss the perspectives for extending the energy 
spectrum analysis to $\tilde{t_2}$ decays and Section 5 has the conclusions.

\section{SUSY Benchmark and Simulation}

The post-{\sc Lep} CMSSM benchmark points E and H, proposed in~\cite{cmssm}, have been 
chosen for this study. 
Point E~\footnote{$m_{1/2}$ = 300, $m_0$ = 1500, $\tan \beta$ = 10, $sign(\mu)$ = +, 
$A$ = 0} 
is representative of the focus region, characterised by heavy sleptons and squarks. 
The $\tilde{\mu_L}^+\tilde{\mu_L}^-$ threshold is just accessible at $\sqrt{s}=$3~TeV.
The gluino  ($M_{\tilde{g}}$=697~GeV) being lighter than the squarks, decays such as 
$\tilde{q} \to q \tilde{g}$ are open for these parameters. It is interesting to observe 
that such inverted scenario, with the gluino lighter than the squarks, is also common 
to benchmarks inspired to effective field theories derived from strings~\cite{kane}.
The mass of the LSP is 120~GeV and those of the heavier neutralinos are in the range 
200-300~GeV. Point H~\footnote{$m_{1/2}$ = 1500, $m_0$ = 419, $\tan \beta$ = 20, 
$sign(\mu)$ = +, $A$ = 0} 
is located at the end-point of the coannihilation tail at large $m_{1/2}$ and 
moderate $\tan \beta$. The heavy sparticle spectrum makes this a highly problematic 
point at the {\sc Lhc}. However, at $\sqrt{s}$=3~TeV, smuons are accessible in $e^+e^-$ 
collisions and their masses can be determined through the muon energy distribution.
The LSP mass, $M_{\chi^0_1}$, is 665~GeV.
Events have been simulated using the {\sc Pythia 6.2} Monte Carlo 
generator. Parameters have been tuned to reproduce sparticles masses and 
widths to a good accuracy. Initial state radiation (ISR) has been included.
Muon energy distributions have been obtained using full {\sc Geant-3} simulation of 
a discrete silicon tracker consisting of twelve layers located from 3.5~cm up to 180~cm 
in radius and a solenoidal magnetic field $B$ of either 4 or 6~T. Backgrounds 
and $\tilde{t}$ decays have been processed using the {\sc Simdet} parametric 
simulation~\cite{simdet}, describing the detector response determined using {\sc Geant}.
The {\sc Clic} luminosity spectrum has been folded to the production cross sections 
using the {\sc Calypso} interface. $\gamma \gamma \rightarrow {\mathrm{hadrons}}$ 
and parallel muon backgrounds have been overlayed to the generated signal and background 
events for the smuon study. 

\section{Muon Mass Determination}

\begin{table}
\caption[]{\sl Main properties of the sparticles considered in this analysis}
\begin{center}
\begin{tabular}{|l|c|c|c|c|c|}
\hline
Point &  & Mass & Width & $\sigma(e^+e^- \rightarrow \tilde{\mu}\tilde{\mu})$ (fb) &
BR($\tilde{\mu} \rightarrow \mu \chi^0_1$) \\
      &  & (GeV) & (GeV) & (fb) &  \\ \hline \hline
E & $\tilde{\mu}_R$ & 1433 & 21.~ & 0.03 & 0.35 \\
E & $\tilde{\mu}_L$ & 1427 & ~7.9 & 0.02 & 0.95 \\ \hline
H & $\tilde{\mu}_R$ & ~710 & ~0.06 & 0.92 & 1.00 \\
H & $\tilde{\mu}_L$ & 1150 & ~0.6 & 0.59  & 1.00 \\ \hline \hline
 &  & &  & $\sigma(e^+e^- \rightarrow \tilde{t} \tilde{t}$) (fb) &
BR($\tilde{t} \rightarrow t \tilde{g}$) \\ \hline
E & $\tilde{t}_1$ & ~994 & 42.~ & 0.65 & 0.20 \\
E & $\tilde{t}_2$ & 1303 & 76.~ & 0.12 & 0.48 \\ \hline
\end{tabular}
\vspace*{-0.5cm}
\end{center}
\label{tab:prop}
\end{table}

The study has been performed for $e^+e^- \to \tilde{\mu_L}\tilde{\mu_L} \to 
\mu^+\chi^0_1 \mu^-\chi^0_1$.
The three main sources of background, also leading to two muons plus missing energy, 
are i) $e^+e^- \rightarrow W^+W^- \rightarrow \mu^+ \mu^- \nu_{\mu} \bar{\nu}_{\mu}$, 
ii) $e^+e^- \to W^+W^- \bar{\nu}\nu \to \mu^+ \mu^- \nu_{\mu} \bar{\nu}_{\mu} \nu_e 
\bar{\nu}_e$ and iii) $e^+e^- \to \chi_1 \chi_2, ~\chi_2 \chi_2 \to 
\mu^+ \mu^- \nu \bar{\nu} \chi_0 \chi_0$. These backgrounds can be suppressed by 
requiring central production and decay kinematics compatible with those characteristic 
of smuon pair production. A multi-dimensional discriminant based on 
$M_{\mu\mu}$, $M_{recoil}$, $E_{missing}$, $\mu \mu$ Acolinearity, 
$|\cos \theta_{Thrust}|$, $E_t$ and $E_{hem}$ has been applied. The signal efficiency is 
flat with the muon energy.

\subsection{The Energy Distribution method}

\begin{figure}[hb!]
\begin{center}
\vspace*{-0.75cm}
\begin{tabular}{c c}
\epsfig{file=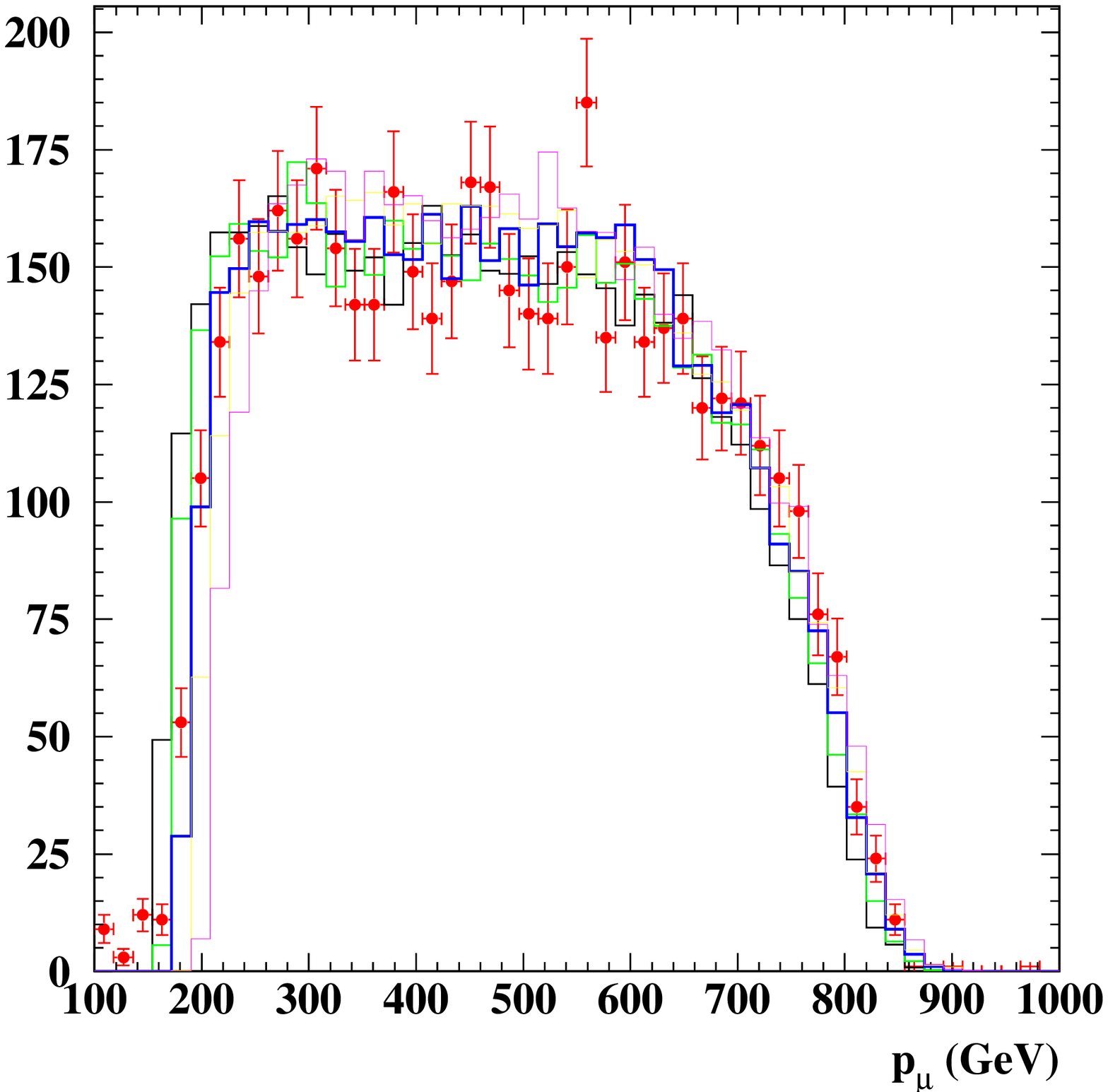,width=6.75cm,height=6.5cm,clip} &
\epsfig{file=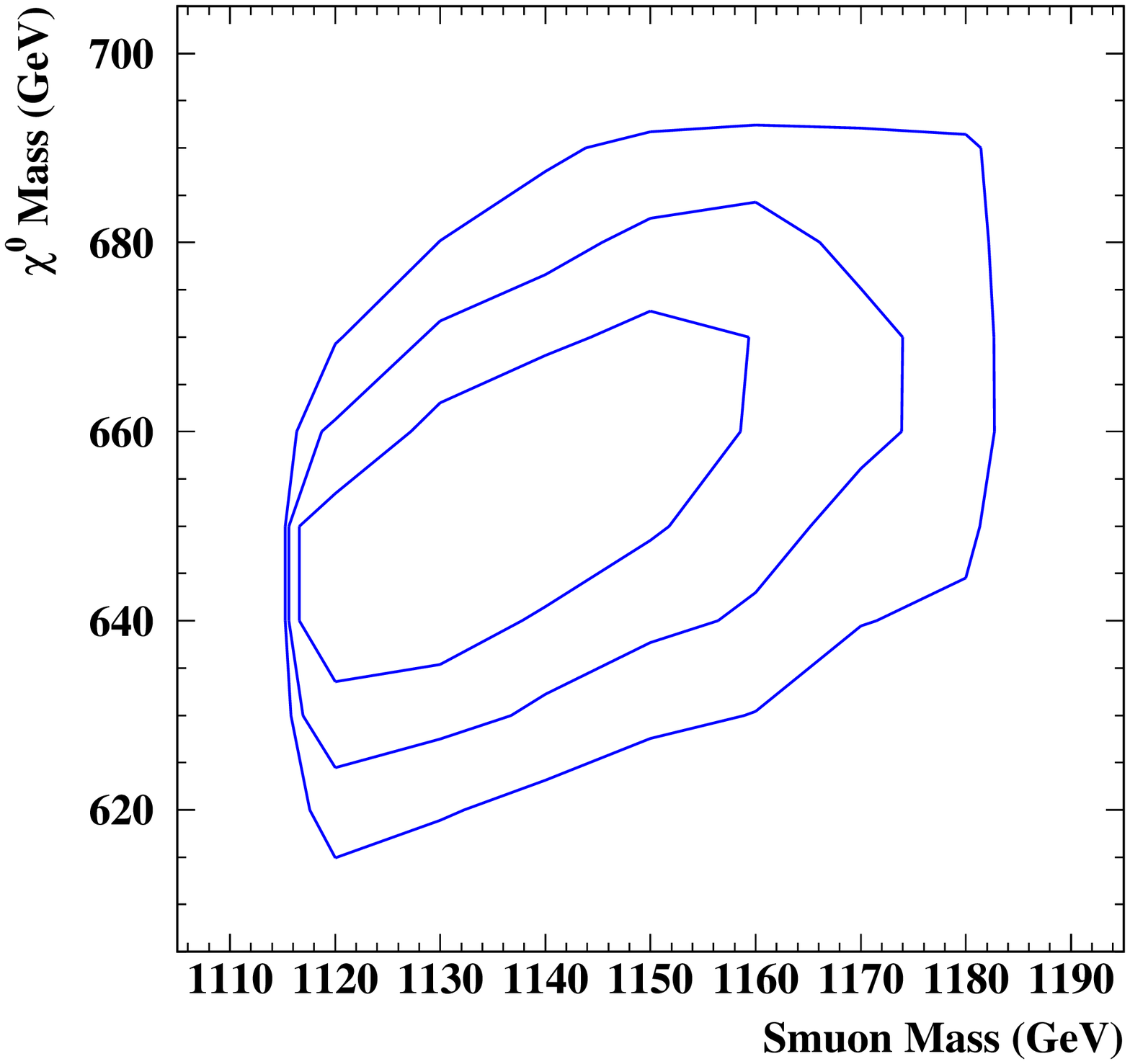,width=6.75cm,height=6.5cm,clip} \\
\end{tabular}
\vspace*{-0.25cm}
\caption{\sl Left: Muon energy spectra in the decay $\tilde{\mu} \rightarrow \mu 
\chi^0_1$ for $M_{\tilde{\mu}}$ = 1150~GeV and $M_{\chi^0_1}$ = 660~GeV obtained at 
$\sqrt{s}$ = 3~TeV, assuming the baseline {\sc Clic} luminosity spectrum.
Right: Accuracy on the determination of the $\tilde{mu}$ and $\chi^0_1$ masses by a 
two-parameter fit to the muon energy distribution. The lines give the contours at 
1~$\sigma$, 68\% and 95\% C.L. for 1~ab$^{-1}$ of data at $\sqrt{s}$=3~TeV.}
\end{center}
\label{fig:pmu}
\end{figure}

If the centre-of-mass energy, $\sqrt{s}$, is significantly larger than twice the 
sparticle mass, $M_{\tilde{\mu}}$, this can be determined by an analysis of the energy 
spectrum of the muon, emitted in the two-body $\tilde{\mu} \rightarrow \chi^0_1 \mu$ 
decay (see Figure~\ref{fig:pmu}).
The two end-points, $E_{min}$ and $E_{max}$, of the spectrum are related to the 
$\tilde{\mu}$ and $\chi^0_1$ masses and to the $\tilde{\mu}$ boost by:
\begin{equation}
E_{max/min} = \frac{M_{\tilde{\mu}}}{2} (1 - \frac{M^2_{\chi^0_1}}{M^2_{\tilde{\mu}}}) 
\times (1 \pm \sqrt{1 - \frac{M^2_{\tilde{\mu}}}{E^2_{beam}}})
\end{equation}
from which either the smuon mass $M_{\tilde{\mu}}$ can be extracted, if $M_{\chi^0_1}$ 
is already known, or both masses can be simultaneously fitted. 
This technique, already considered for the determination of squark masses~\cite{feng}, 
has been extended also 
to sleptons for the {\sc Lhc} and a TeV-class LC~\cite{martyn}. It is interesting to 
consider it here also for its implications on the required momentum resolution in the 
detector. Two values of the solenoidal magnetic field $B =$ 4 and 6~T have been tested, 
corresponding to momentum resolutions $\delta p/p^2$ of $4.5 \times 10^{-5}$ and 
$3.0 \times 10^{-5}$~GeV$^{-1}$ respectively. No appreciable difference on the 
resulting mass accuracy has been observed from these two momentum resolution, since 
the dominant smearing is due to beamstrahlung.
In fact, at {\sc Clic}, the main issue is the significant beamstrahlung smearing of the 
luminosity spectrum, and thus of the effective $E_{beam}$ value. The corresponding  
effect has been estimated by using both a perfectly well known and constant beam energy 
and that corresponding to the baseline {\sc Clic} parameters at a nominal 
$\sqrt{s}$=3~TeV. Results are summarised in Table~\ref{tab:res1}.
\begin{table}
\begin{center}
\caption{\sl Results of the 1-parameter $\chi^2$ fit to the muon energy distribution, 
obtained for different assumption on the $\delta p/p^2$ momentum resolution and the 
beamstrahlung spectrum. Accuracies are given for an integrated luminosity of 
1~ab$^{-1}$.}
\vspace*{0.25cm}  
\begin{tabular}{|c|c|c|}
\hline
$\delta p/p^2$ & Beamstrahlung & Fit Result (GeV) \\ \hline \hline
0. & none & 1150 $\pm$ 10 \\ \hline 
3.0 $\times 10^{-5}$ & none & 1150 $\pm$ 12 \\
4.5 $\times 10^{-5}$ & none & 1151 $\pm$ 12 \\ \hline
4.5 $\times 10^{-5}$ & Std. & 1143 $\pm$ 18 \\ \hline
\end{tabular}
\vspace*{-0.5cm}
\end{center}
\label{tab:res1}
\end{table}
The smuon mass has been extracted by a $\chi^2$ fit to the muon energy spectrum by 
fixing $M_{\chi^0_1}$ to its nominal value (see Table~\ref{tab:res1}. 
The fit has been also repeated leaving both masses free and 
performing a simultaneous two-parameter fit. The results are $M_{\tilde{\mu}} = 
(1145 \pm 25)$~GeV and $M_{\chi^{0}_1} = (652 \pm 22)$~GeV (see Figure~\ref{fig:pmu}).

\subsection{The Threshold Scan method}

An alternative method to determine the $\tilde{\mu}$ mass is an energy scan of the rise 
of the $e^+e^- \rightarrow \tilde{\mu}^+ \tilde{\mu}^-$ cross section, close to its
kinematical threshold. It has been shown that an optimal scan consists of just two 
energy points, sharing the total integrated luminosity in equal fractions and chosen 
at locations optimising the sensitivities to the $\tilde{\mu}$ width and mass 
respectively~\cite{blair}. Including the beamstrahlung effect, induces a shift of the 
positions of the maxima in mass sensitivity towards higher nominal $\sqrt{s}$ energies.
For benchmark point E, the cross section at $\sqrt{s}$=3~TeV is too small for an 
accurate measurement. Higher centre-of-mass energy, 4~TeV, and polarised beams need to 
be considered. By properly choosing the beam polarisation, not only the pair production 
cross sections are increased but also their sensitivity to the smuon masses. Results 
are summarised in Table~\ref{tab:res2}.

\begin{table}
\caption[]{\sl Accuracies on the determinations of the smuon masses for the two 
benchmark scenarios considered in this study using threshold scans with different 
experimental conditions.}
\begin{center}
\begin{tabular}{|l|c|c|c|c|c|c|c|}
\hline
Point & &Beam- & Pol. & $\sqrt{s}$ & $\int{\cal{L}}$ & $\delta M$ \\
      & &strahlung &  &  (TeV) & (ab$^{-1}$)          & (GeV) \\ \hline \hline
H     & $\tilde{\mu}_L$ & none & 0/0 & 3.0-3.5   & 1 & $\pm$ 11 \\
H     & $\tilde{\mu}_L$ & Std. & 0/0 & 3.0-3.5   & 1 & $\pm$ 15 \\ \hline
E     & $\tilde{\mu}_L$ & none & 0/0 & 3.8-4.2 & 1 & $\pm$ 29 \\ 
E     & $\tilde{\mu}_L$ & Std. & 0/0 & 3.8-4.2 & 1 & $\pm$ 36 \\ 
E     & $\tilde{\mu}_L$ & none & 80/60 & 3.8-4.2 & 1 & $\pm$ 17 \\ 
E     & $\tilde{\mu}_L$ & Std. & 80/60 & 3.8-4.2 & 1 & $\pm$ 22 \\ \hline
\end{tabular}
\vspace*{-0.5cm}
\end{center}
\label{tab:res2}
\end{table}

\section{Scalar Top Analysis}

\begin{figure}[h!]
\vspace*{-0.75cm}
\begin{center}
\centerline{\epsfig{file=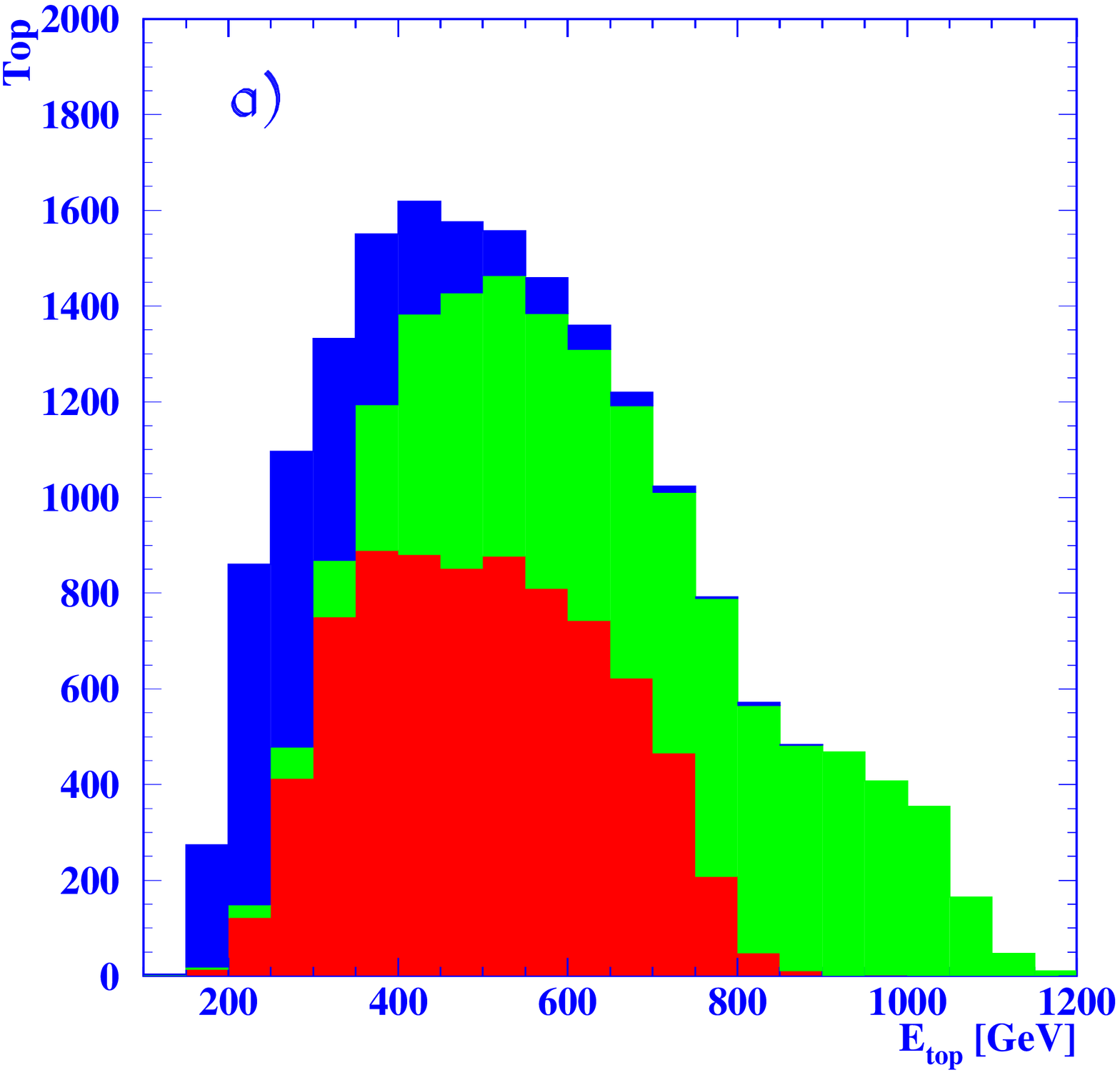,height=8.0cm,width=6.95cm} 
            \epsfig{file=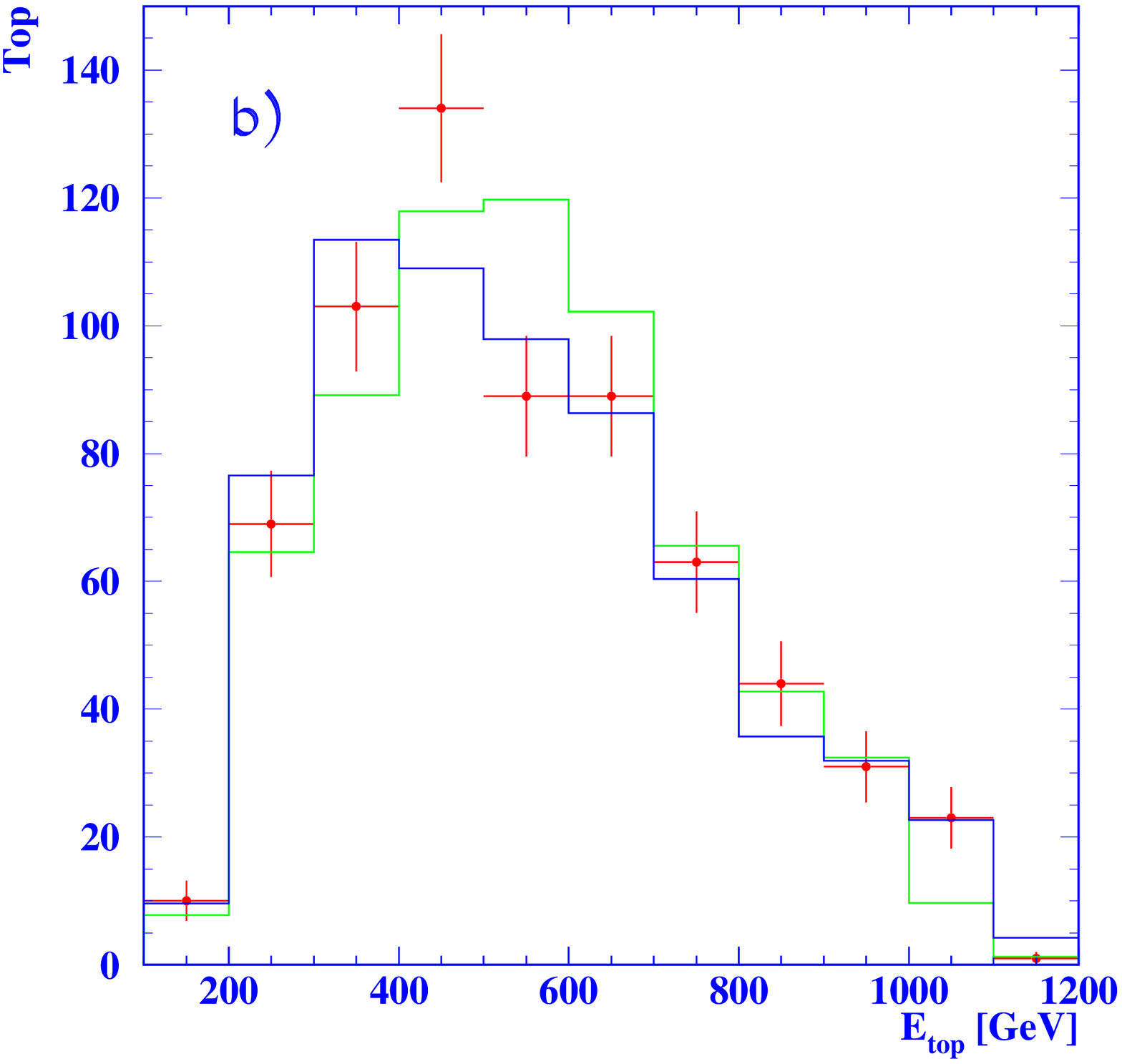,height=8.0cm,width=6.95cm}}
\vspace*{-0.25cm}
\caption{\sl Energy spectrum of top quarks in  $e^+e^- \rightarrow 
\tilde{t_2} \tilde{t_2}$ events. The $\tilde{t_2} \to t \tilde{g}$ signal is
represented by the lower histogram. 
Left: the different contributions to the spectrum. Right: $E_{top}$ spectrum 
corresponding to the nominal mass and 3~ab$^{-1}$ (points with error bars) compared 
to the expectations for $M_{\tilde{t_2}}$ = 1300$\pm$60~GeV. The endpoint from 
$\tilde{t_2} \rightarrow t  \tilde{g}$ remains visible when overlayed to the 
distributions from the other decay channels.}
\end{center}
\vspace*{-0.5cm}
\label{fig:stop}
\end{figure}

In the context of the benchmark point E, the process 
$e^+e^- \rightarrow \tilde{t_1} \tilde{t_1}$, $\tilde{t_2} \tilde{t_2}$ followed by the 
decay $\tilde{t_{1.2}} \rightarrow t  \tilde{g}$ offers a possibility to determine the 
scalar top mass based on the reconstructed top energy only, following a procedure 
similar to that adopted for the smuon mass.
The same relation as Eq.~1 between the top energy spectrum endpoints and the masses 
and $\sqrt{s}$ energy holds in this case, with $M_{\tilde{t}}$ and $M_{\tilde{g}}$ 
replacing $M_{\tilde{\mu}}$ and $M_{\chi^0_1}$ respectively. By determining the 
endpoints of the top quark energy spectrum produced in 
$\tilde{t} \rightarrow t \tilde{g}$, the masses of the stop and the gluino can be 
extracted. However, here the situation is complicated by the fact that top quarks may 
also be produced in $\tilde{g}$ decays or other $\tilde{t}$ decays, such as 
$\tilde{t} \rightarrow t  \chi^0$. 
Figure~\ref{fig:stop} shows the distribution of the energy of top quarks produced 
in $e^+e^- \rightarrow \tilde{t_2} \tilde{t_2}$ events. 
Although the expected endpoints are diluted, it is still possible to extract 
$M_{\tilde{t}}$ from the shape of the top energy spectrum.
A $\chi^2$ fit, including the contributions from other top sources in 
signal events, indicates that a relative accuracy 
$\delta M_{t_{2}}/M_{t_{2}} = \pm$7.5\% can be achieved with 3~ab$^{-1}$, without 
including beamstrahlung effects. 

\section{Conclusions}

A preliminary study of sfermion reconstruction at the {\sc Clic} multi-TeV collider
has shown that smuon masses can be measured with typical accuracies of 1.3-3.0~\% for 
${\cal{L}}=1$~ab$^{-1}$ for two benchmark points, using either the muon energy 
technique, when $\sqrt{s} >> 2 \times M_{\tilde{\mu}}$, or a threshold scan. 
The accuracy of the threshold scans is significantly enhanced by the availability of 
polarised beams.
The energy spectrum technique has also been extended to the scalar top in the case when 
the decay $\tilde{t_2} \to t \tilde{g}$ is kinematically allowed. 

\vspace*{0.5cm}

{\sl We are grateful to G.~Blair, A.~De~Roeck, J.~Ellis, S.~Kraml and L.~Pape for 
contributions.}

\end{document}